\documentclass[aip,jap,reprint,10pt,showpacs,altaffillsymbol,superscriptaddress,longbibliography]{revtex4-1}
\pdfoutput=1
\usepackage{graphicx}
\usepackage{amsfonts}
\usepackage{amsmath}
\usepackage{natbib}

\begin{document}

\title{Re-evaluation of uncertainty for calibration of $100$~M$\Omega$ and $1$~G$\Omega$ resistors at NPL}

\author{S.~P.~Giblin}
\affiliation{National Physical Laboratory, Hampton Road, Teddington, Middlesex TW11 0LW, United Kingdom}

\email[stephen.giblin@npl.co.uk]{Your e-mail address}

\date{\today}

\begin{abstract}
The uncertainty for calibrating  $100$~M$\Omega$ and $1$~G$\Omega$ resistors using the NPL high-resistance CCC bridge has been re-evaluated, resulting in combined uncertainties of $\sim 0.1$~$\mu\Omega/ \Omega$, considerably smaller than the CMC entries of $0.4$~$\mu\Omega/ \Omega$ and $1.6$~$\mu\Omega/ \Omega$ respectively. This re-evaluation supports measurement of single electron pump currents with combined uncertainties of a few parts in $10^7$.
\end{abstract}

\pacs{1234}

\maketitle

\section{Introduction}
Traceable measurements of high resistance are required for diverse applications including semiconductor wafer characterization, insulator testing and radiation dosimetry. Recently, research into semiconductor electron pumps\cite{kaestner2015non}, motivated by the upcoming re-definition of the SI base unit ampere\cite{pekola2013single}, has focused additional attention onto high resistance and small current metrology. Semiconductor electron pumps are micron-scale, cryogenic devices, which generate a current by transporting electrons one at a time in response to a clock input with frequency $f$, yielding a current ideally equal to $ef$, where e is the elementary charge. Promising early experiments using conventional laboratory ammeters to measure the current showed that these pumps could generate quantized current plateaus with $f$ of order $\sim 1$~GHz, generating pump currents $I_{\text{P}}$ larger than $100$~pA \cite{blumenthal2007gigahertz,fujiwara2008nanoampere}. Since then, metrological research has focused on measuring $I_{\text{P}}$ as accurately as possible to investigate possible error processes, and so far seven studies have shown the pump current to be accurately quantised within measurement uncertainties of $1$ part-per-million or less \cite{giblin2012towards, bae2015precision, stein2015validation, yamahata2016gigahertz, stein2016robustness, giblin2017robust, zhao2017thermal}. Five of these studies\cite{giblin2012towards, bae2015precision, yamahata2016gigahertz, giblin2017robust, zhao2017thermal} were performed at the National Physical Laboratory (NPL), UK, using a measurement system which compared $I_{\text{P}}$ to a reference current $I_{\text{R}}$ derived from a measured voltage across a calibrated $1$~G$\Omega$ standard resistor\cite{giblin2012towards}. The focus of this paper is a precise evaluation of the uncertainty in calibrating this resistor.

The high-resistance cryogenic current comparator (CCC)\cite{fletcher2000cryogenic} has been in routine use at NPL since 2001 for calibration of decade-value standard resistors in the range from $100$~k$\Omega$ to $1$~G$\Omega$. The calibration and measurement capability (CMC) expanded relative uncertainties ($k=2$) are $0.08\times 10^{-6}$, $0.12\times 10^{-6}$, $0.2\times 10^{-6}$, $0.4\times 10^{-6}$ and $1.6\times 10^{-6}$ for decade standard resistors of nominal value $100$~k$\Omega$, $1$~M$\Omega$, $10$~M$\Omega$, $100$~M$\Omega$ and $1$~G$\Omega$ respectively. For the first four of the NPL electron pump measurement campaigns\cite{giblin2012towards, bae2015precision, yamahata2016gigahertz, giblin2017robust}, the type B uncertainty in the $1$~G$\Omega$ resistor value was simply assigned to be the CMC uncertainty, giving a $1\sigma$ relative uncertainty contribution of $8 \times 10^{-7}$. This was the largest single contribution to the uncertainty in $I_{\text{P}}$ in all these studies. The relative uncertainty in measuring the voltage across the resistor was reduced to $\sim 1 \times 10^{-7}$ by frequently calibrating the voltmeter directly against a Josephson voltage standard\cite{giblin2014sub}. Averaging up to two days of data\cite{giblin2017robust} reduced the type A uncertainty in $I_{\text{P}}$ to a similar level, leaving the $1$~G$\Omega$ uncertainty completely dominating the uncertainty budget. Thus, the present study was undertaken to re-evaluate the type B uncertainty in the $1$~G$\Omega$ calibration. The re-evaluated uncertainty was used in a recent measurement campaign on a silicon electron pump, yielding a relative combined uncertainty of $2.7 \times 10^{-7}$ for measuring a pump current of $\sim 160$~pA\cite{zhao2017thermal}.

An alternative approach to reducing the uncertainty could have been to use a lower-value resistor to generate $I_{\text{R}}$, for example $100$~M$\Omega$ instead of $1$~G$\Omega$. This involves a trade-off: lower type B uncertainty is attained at the expense of larger type A uncertainty because Johnson-Nyquist current noise scales as $1/ \sqrt{R}$. For the NPL CMC calibration uncertainties, the trade-off becomes favorable for $I_{\text{P}} \gtrsim 150$~pA\cite{giblin2014sub}. However, recent investigations have shown that available $100$~M$\Omega$ and $1$~G$\Omega$ standard resistors based on thick-film elements may not be sufficiently stable, on time-scales of hours to days, to routinely achieve relative uncertainties much less than $\sim 5\times 10^{-7}$ in practice\cite{giblin2018limitations}. Much lower-value resistors are required to achieve sufficient stability, but as noted, these have unacceptably high thermal current noise. One innovative approach is to combine a $1$~M$\Omega$ current-to-voltage conversion resistor with a stable $1000:1$ current-scaling resistor network. This package, known as the ultrastable low-noise current amplifier (ULCA)\cite{drung2015ultrastable}, combines the stability and low relative calibration uncertainty ($<1 \times 10^{-7}$) of the $1$~M$\Omega$ resistor with the low thermal noise of the current scaling network, which (in the standard ULCA) presents a $3$~G$\Omega$ resistance to the input. The ULCA has been used at the Physikalisch-Technische Bundesanstalt (PTB)\cite{stein2015validation,stein2016robustness}, to measure electron pump currents with relative combined uncertainties as low as $1.6 \times 10^{-7}$.

This paper is structured as follows: In section II, we give an overview of the CCC bridge, and describe its mode of operation for calibrating $100$~M$\Omega$ and $1$~G$\Omega$ resistors. In section III we discuss the noise and type A uncertainty. In section IV we review independent tests of the bridge prior to this study. Section V forms the main body of the paper, in which we describe in detail our evaluation of the main sources of bridge error: Leakage resistance, SQUID non-linearity, series resistance correction and winding ratio error. In section VI we discuss further errors affecting the uncertainty in the unknown resistor. We aim to restrict our discussion to the CCC bridge itself, but inevitably in this section we briefly discuss the stability of standard resistors in use at NPL. Finally in section VII we present the full uncertainty budget and discuss the possibilities for lower-noise measurements of electron pump currents.

\begin{figure}
\includegraphics[width=8.5cm]{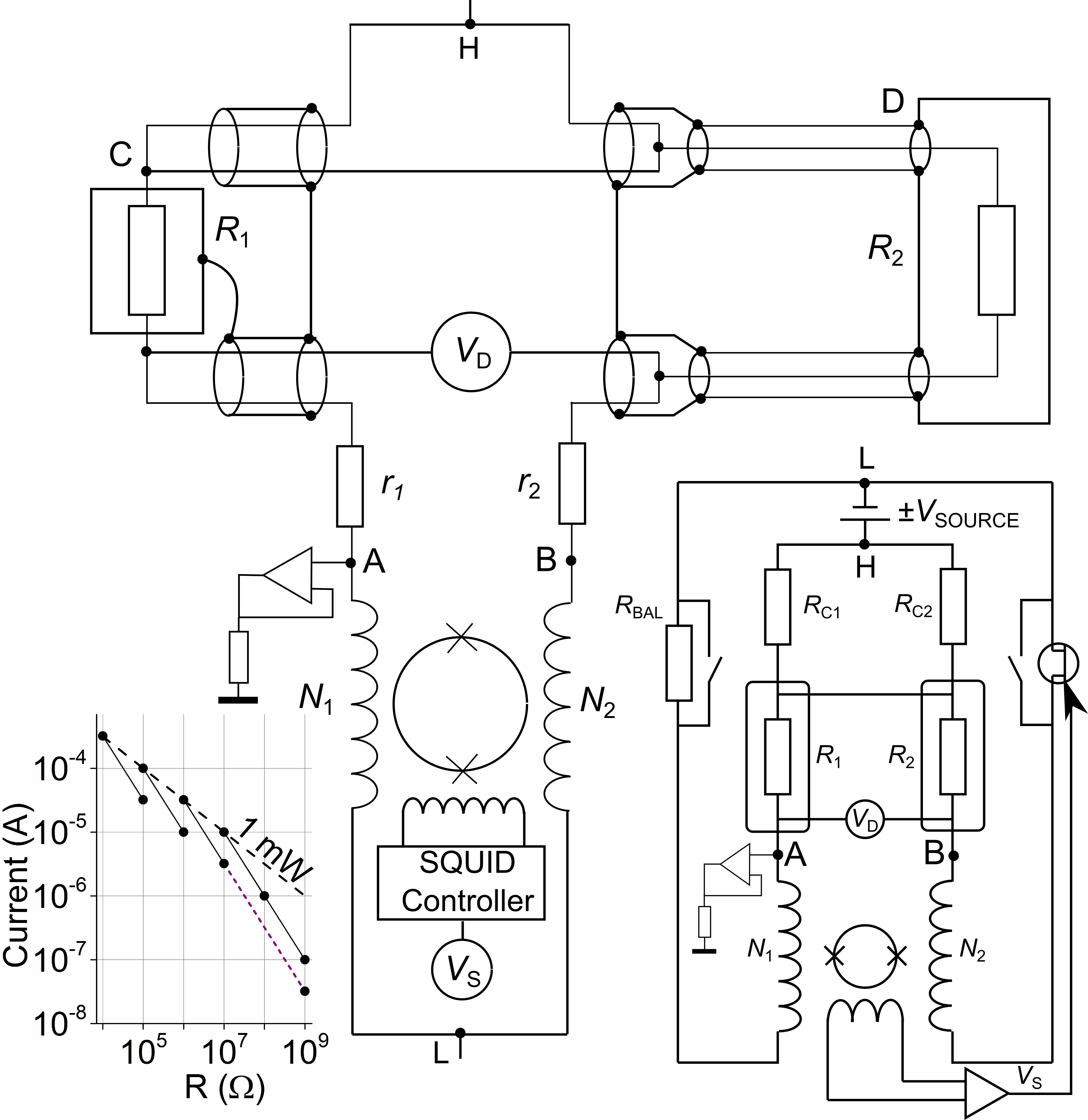}
\caption{\label{BridgeDiagramFig}\textsf{Schematic diagrams of the bridge. The lower right inset shows the general schematic circuit of the bridge, and the main figure shows the detailed connections specific to the case where $R_{\text{1}}=10$~M$\Omega$ and $R_{\text{2}}=100$~M$\Omega$ or $1$~G$\Omega$. The screen is not shown. Points H and L indicate the high and low potential terminals of the voltage source, and points A, B, C and D are referenced in the main text. The lead resistances $r_{\text{1}}$ and $r_{\text{2}}$ are distributed between the junction L and the low terminals of $R_{\text{1}}$ and $R_{\text{2}}$. The lower left inset illustrates the current scaling used in build-up measurements. Solid lines: routine traceability. Dashed line: SET $1$~G$\Omega$ measurement. The dashed line with gradient $-\frac{1}{2}$ shows 1 mW power dissipation}}
\end{figure}

\begin{figure}
\includegraphics[width=8.5cm]{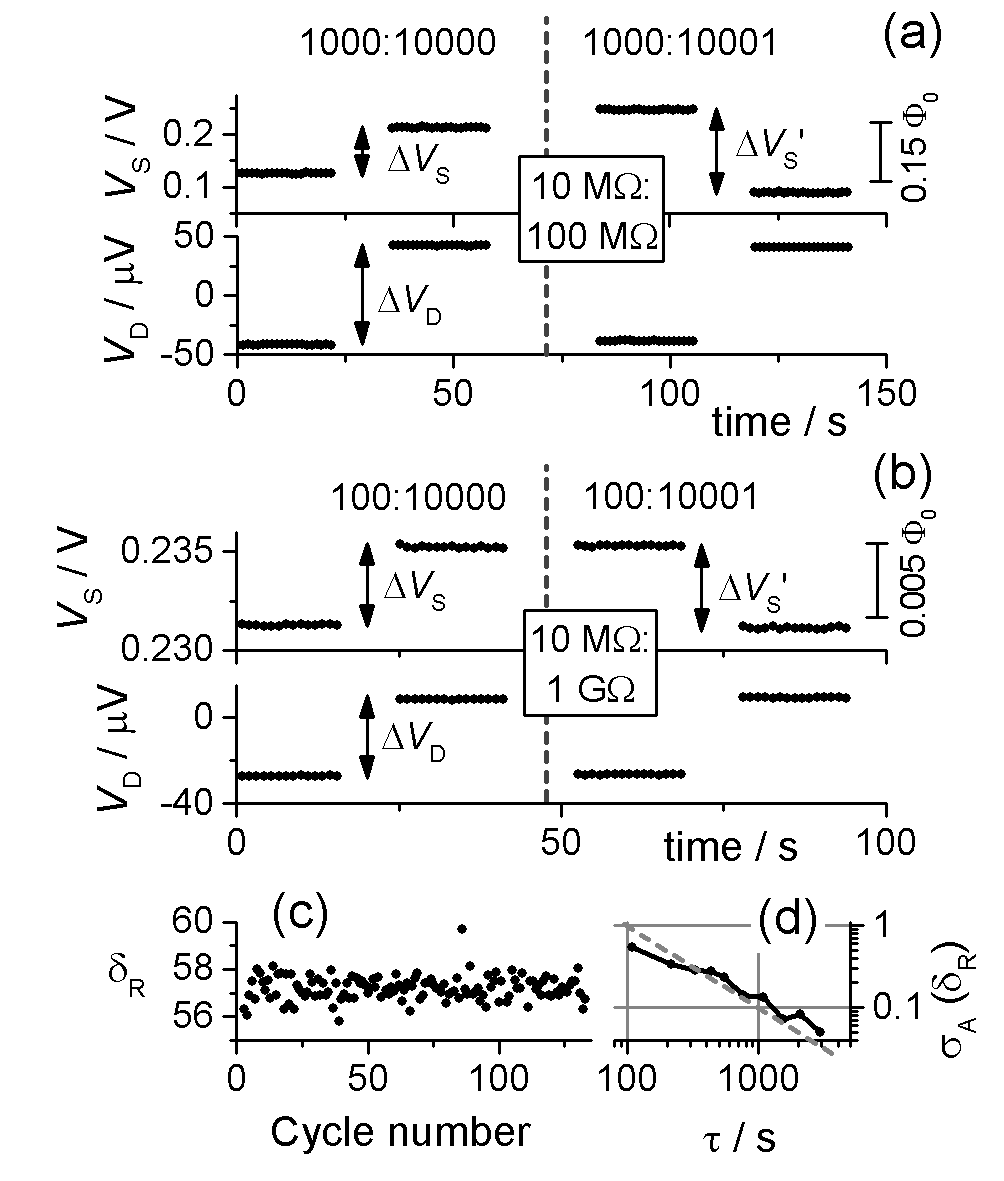}
\caption{\label{RawDataFig}\textsf{Examples of raw SQUID and detector data for one measurement cycle. (a): $10$~M$\Omega$:$100$~M$\Omega$ with $V_{\text{source}}=\pm100$~V. (b): $10$~M$\Omega$:$1$~G$\Omega$ (the SET $1$~G$\Omega$) with $V_{\text{source}}=\pm33$~V. Vertical arrows indicate the difference voltages $\Delta V_{\text{S}}$, $\Delta V_{\text{S}}'$ and $\Delta V_{\text{D}}$ used to calculate the resistance ratio using equation (4) of the main text. Vertical dashed lines separate the two phases of the cycle with $N_{\text{2}}=10000$ (left side of the line) and $N_{\text{2}}=10001$ (right side of the line). Solid vertical bars provide a SQUID flux scale. (c): Example measurement of the SET $1$~G$\Omega$: $R_{\text{1G}}=1$~G$\Omega(1+\delta_{\text{R}}/10^6)$. Each data point is calculated from one cycle of the type shown in panel (b). (d): Allan deviation of the data shown in (c). The dashed line is a guide to the eye with slope $1/\sqrt{\tau}$}}
\end{figure}

\section{bridge circuit and operation}
A schematic diagram of the overall bridge circuit is shown in the lower right inset to figure~\ref{BridgeDiagramFig}. The design departs from more conventional CCC resistance bridges, which employ two isolated current sources to drive current in the two arms of the bridge\cite{williams1991automated,gotz2009improved,williams2010automated}. Our high-resistance CCC employs a single-source topology, which places less stringent demands on the noise properties of the source\cite{fletcher2000cryogenic,hernandez2014precision}. In our circuit, the source voltage $V_{\text{SOURCE}}$ drives current in the two resistors to be compared, $R_{\text{1}}$ and $R_{\text{2}}$, and the CCC windings with $N_{\text{1}}$ and $N_{\text{2}}$ turns. Relays allow selection of $N_{\text{1}}=100,1000,10000$. $N_{\text{2}}$ is fixed at $10000$ turns, with an additional turn added periodically by a relay, as will be detailed below. As in more conventional dual-source CCC bridges \cite{williams1991automated,williams2010automated}, a Wagner circuit maintains the junction of $R_{\text{1}}$ and $N_{\text{1}}$ (point A in figure \ref{BridgeDiagramFig}) at screen potential to minimise errors due to leakage resistance. A combining network composed of resistors $R_{\text{C1}}$ and $R_{\text{C2}}$ reduces to a negligible level the effect of current flowing in the high potential terminals of $R_{\text{1}}$ and $R_{\text{2}}$, but this plays no role for the high value, two-terminal resistors mostly considered in this paper. 

The routine build-up procedure (lower left inset to figure \ref{BridgeDiagramFig}) involves a series of 10:1 (or 100:1 for measuring $1$~G$\Omega$) measurements with the lower-value resistor dissipating $1$~mW of power. Measurements of $1$~G$\Omega$ for the single-electron current source used a lower value of $V_{\text{SOURCE}} = \pm 33$~V instead of $\pm 100$~V for two reasons: firstly, to eliminate any possible voltage co-efficient in the $10$~M$\Omega$ reference resistor (since the $1$~M$\Omega$ : $10$~M$\Omega$ and $10$~M$\Omega$ : $1$~G$\Omega$ measurements are made at the same voltage), and secondly, to reduce the possible voltage co-efficient in the $1$~G$\Omega$, a point discussed further in section VI.B.

For $[R_{\text{1}}:R_{\text{2}}] = [10$~k$\Omega:100$~k$\Omega]$ up to [$1$~M$\Omega:10$~M$\Omega]$, the squid output $V_{\text{S}}$ is used as the input to a digital servo, which controls the gate voltage of a JFET in one of the bridge arms. This controls the resistance of the JFET channel, and ensures the current balance condition $I_{\text{1}}/I_{\text{2}} = N_{\text{2}}/N_{\text{1}}$. The JFET resistance is balanced by a fixed resistor $R_{\text{BAL}}$ in the other bridge arm. However, for the measurements we are mostly concerned with in this paper, $[R_{\text{1}}:R_{\text{2}}] = [10$~M$\Omega:100$~M$\Omega]$ and [$10$~M$\Omega:1$~G$\Omega]$, the servo is not used and the bridge functions as a passive current divider in which $I_{\text{1}}/I_{\text{2}} = R_{\text{B}}/R_{\text{A}}$ with $R_{\text{A}} = R_{\text{1}}+r_{\text{1}}$ and $R_{\text{B}} = R_{\text{2}}+r_{\text{2}}$. The series resistances $r_{\text{1}}$ and $r_{\text{2}}$ (main panel of figure \ref{BridgeDiagramFig}) are due to wiring in the room-temperature electronics, connecting cables and the down-leads in the CCC cryostat.

\begin{figure}
\includegraphics[width=8.5cm]{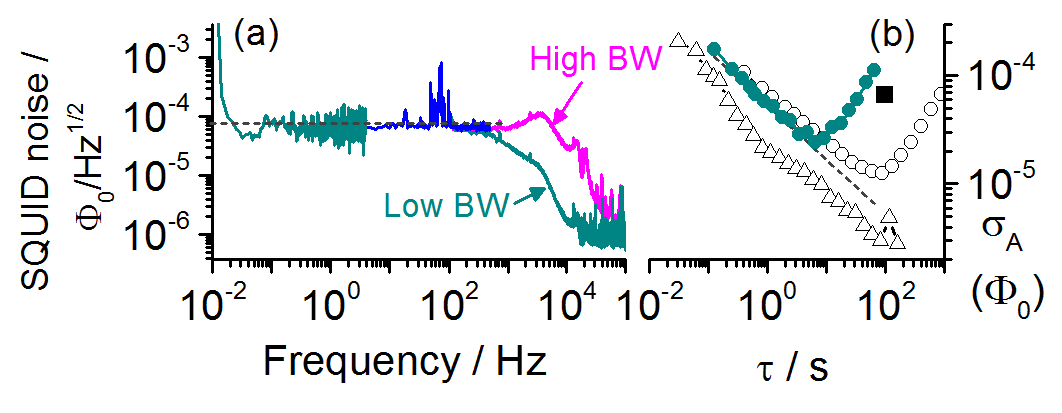}
\caption{\label{NoiseFig}\textsf{(a): Amplitude spectrum of the voltage output of the SQUID control unit, referred to flux noise using the conversion factor $0.75$~V/$\Phi_0$. The high frequency data was acquired using a $16$-bit digitiser sampling at $500$~ksamples/s, and the low frequency data was obtained by block-averaging $50000$ samples from the digitiser. The intermediate frequency range was covered using an integrating voltmeter with an aperture of $400$~$\mu$s. High frequency data was obtained with the SQUID controller feedback loop set to high- and low-bandwidth modes. (b): filled circles: Allan deviation calculated from the same time-domain data set that yielded the low-frequency portion of the spectrum in plot (a). Open circles: Allan deviation calculated from a data set obtained $7$ years prior to the data in this study. Open triangles: Allan deviation calculated from a data set obtained $7$ days after the data in panel (a). The single filled square shows the data point from figure \ref{RawDataFig} (d) at $\tau = 100$~s converted to flux units. The dashed lines in both plots show a white noise level of $75$~$\mu \Phi_0 / \sqrt{Hz}$.}}
\end{figure}

The SQUID output is proportional to the difference of the ampere-turns product in the two windings: $V_{\text{S}} = k(I_{\text{1}}N_{\text{1}} - I_{\text{2}}N_{\text{2}})$. The sensitivity co-efficient $k$  is given by $k=S_{\text{SQUID}} / S_{\text{CCC}}$ with $S_{\text{SQUID}} \approx 0.7$~V$/ \Phi_{\text{0}}$ and $S_{\text{CCC}} \approx 6$~$\mu$A turns$/ \Phi_{\text{0}}$. To eliminate $k$ from the bridge equations, the measurement is performed in two phases, with $N_{\text{2}} = 10000$ and  $N_{\text{2}}' = 10001$. For each phase, the voltage source is reversed, yielding SQUID difference signals $\Delta V_{\text{S}}$ and $\Delta V_{\text{S}}'$. The detector voltage difference $\Delta V_{\text{D}}$ from the first phase is also extracted from the data. Examples of raw data for  $100$~M$\Omega$ and  $1$~G$\Omega$ measurements are shown in figures~\ref{RawDataFig} (a) and (b) respectively. The resistance ratio in the two arms of the bridge is then given by 

\begin{equation}
\frac{R_{\text{B}}}{R_{\text{A}}} = \frac{N_{\text{2}}' \Delta V_{\text{S}} - N_{\text{2}} \Delta V_{\text{S}}'}{N_{\text{1}}(\Delta V_{\text{S}}-\Delta V_{\text{S}}')}.
\end{equation}

Since $r_{\text{1}}\ll R_{\text{1}}$, $r_{\text{2}}\ll R_{\text{2}}$, we can re-write equation (1) as

\begin{equation}
\frac{R_{\text{2}}}{R_{\text{1}}} = \left( \frac{N_{\text{2}}' \Delta V_{\text{S}} - N_{\text{2}} \Delta V_{\text{S}}'}{N_{\text{1}}(\Delta V_{\text{S}}-\Delta V_{\text{S}}')} \right) \left( 1+\frac{r_{\text{1}}}{R_{\text{1}}}-\frac{r_{\text{2}}}{R_{\text{2}}}\right)
\end{equation}

It is necessary to measure $r_{\text{1}}$ with an uncertainty less than $0.1$~$\Omega$ in order for this correction to contribute less than $1 \times10^{-8}$ to the relative uncertainty in $R_{\text{2}}$. The series resistances can be measured directly with a digital multimeter\cite{bierzychudek2009uncertainty}, and we found $r_{\text{1}}\approx 5.3$~$\Omega$ and $r_{\text{2}} \approx 15$~$\Omega$, but for our bridge it is more practical to use the bridge voltage detector signal $V_{\text{D}} = I_{\text{1}}r_{\text{1}} - I_{\text{2}}r_{\text{2}}$. Again, assuming $r_{\text{1}}\ll R_{\text{1}}$, $r_{\text{2}}\ll R_{\text{2}}$, the detector voltage difference $\Delta V_{\text{D}}$ when the source voltage is reversed is given by

\begin{equation}
\Delta V_{\text{D}} = 2V_{\text{SOURCE}} \left( \frac{r_{\text{1}}}{R_{\text{1}}} - \frac{r_{\text{2}}}{R_{\text{2}}} \right).
\end{equation}

Substituting equation (3) into equation (2) yields the final equation for determining the resistance ratio,

\begin{equation}
\frac{R_{\text{2}}}{R_{\text{1}}} = \left( \frac{N_{\text{2}}' \Delta V_{\text{S}} - N_{\text{2}} \Delta V_{\text{S}}'}{N_{\text{1}}(\Delta V_{\text{S}}-\Delta V_{\text{S}}')} \right) \left( 1+\frac{\Delta V_{\text{D}}}{2V_{\text{SOURCE}}}\right).
\end{equation}

The main panel of figure~\ref{BridgeDiagramFig} is a detailed schematic showing the resistor connections to the bridge for the case of $[10$~M$\Omega:100$~M$\Omega]$ and [$10$~M$\Omega:1$~G$\Omega]$ measurements. The $10$~M$\Omega$ resistor $R_{\text{1}}$ is a 2-terminal standard (usually Fluke type 742A) in a temperature-controlled air bath, connected to the CCC with a 4-wire cable consisting of two individually screened twisted pairs. The $100$~M$\Omega$ or $1$~G$\Omega$ resistor $R_{\text{2}}$ usually has two coaxial terminals. The $1$~G$\Omega$ resistor used for the single-electron measurements (of type Guildline 9336) has a measured temperature co-efficient of $5.2 \times 10^{-6} / ^{\circ}$C, and is in a custom-made temperature controlled enclosure with $\pm 0.005 ^{\circ}$C stability. For the remainder of this paper we refer to this artifact as the `SET $1$~G$\Omega$'.

\section{Noise and type A uncertainty}

Figure \ref{NoiseFig} presents frequency-domain (panel(a)) and Allan deviation plots (panel (b)) of the SQUID flux noise, obtained from time traces of $V_{\text{S}}$ measured with the CCC windings disconnected from the electronics (very similar plots were obtained with the electronics connected). Several peaks are visible in the amplitude spectrum, the most prominent of which is a resonance at $80$~Hz which is probably of mechanical origin. The damped L-C winding resonance is at $\sim 4$~kHz. The white noise level of $7.5 \times 10^{-5} \Phi_{0} / \sqrt{Hz}$ (dashed line in both plots) is roughly $20$ times the bare SQUID noise. From the Allan deviation plot (filled circles), a transition from white noise to random walk behavior occurs for times longer than around 10 seconds. However, we also show Allan deviation plots (open triangles and circles) obtained at different times, which demonstrate that the low frequency SQUID noise behaviour exhibits considerable variation in time. A calibration of $1$~G$\Omega$ with $\pm 33$~nA in the $10000$~turn winding presents a full-signal flux linkage of $2I_{\text{1}}N_{\text{1}} / S_{\text{CCC}} = 110$~$\Phi_{0}$. Therefore, based on the measured white flux noise of $7.5 \times 10^{-5} \Phi_{0} / \sqrt{Hz}$, we expect to resolve $7 \times 10^{-8}$ of full signal at the measurement cycle time of $10^{2}$~s. Crucially, the measured noise at $\tau = 100$~s plotted in Fig. \ref{NoiseFig} (b) varies over almost two orders of magnitude and corresponds to relative uncertainties in a $1$~G$\Omega$ measurement from $4 \times 10^{-8}$ to $1.5 \times 10^{-6}$. Results of a measurement of the SET $1$~G$\Omega$ are shown in figure \ref{RawDataFig} (c). This measurement consisted of $128$ cycles, lasting a total $3.5$ hours. The Allan deviation (figure \ref{RawDataFig} (d)) at $\tau = 100$~s is $5.9 \times 10^{-7}$, and this value, converted to flux, is plotted as the single square data point in \ref{NoiseFig} (b), showing that it is consistent with the more pessimistic SQUID noise data. It is immediately apparent that a better control of the low frequency SQUID noise behaviour, combined with an optimisation of the measurement cycle time, could lead to a much lower type A uncertainty for high resistance measurements. 

The Allan deviation plot of figure \ref{RawDataFig} (d) suggests that the resistor is stable on the time-scale of $\sim 1$ hour probed by this relatively short measurement. Extrapolating to longer measurement times, we would expect type A uncertainties of $\sim 2 \times 10^{-8}$ to be achieved with an overnight measurement. In fact, long measurements of the SET $1$~G$\Omega$ and also a sample of $100$~M$\Omega$ standard resistors typically show instability on time-scales longer than a few hours\cite{giblin2018limitations} and consequently, the standard error on the mean is not a meaningful measure of the type A uncertainty\cite{allan1987should}. For the recent high-precision electron pump measurements\cite{zhao2017thermal}, relative type A uncertainties $\sim 1 \times 10^{-7}$ were evaluated on a case-by-case basis by examining the Allan deviation of resistor calibrations performed immediately before and after the pump measurements\cite{zhao2017thermalsupplement}. The main focus of this paper is the uncertainty in the CCC measurement, and the limitations to the standard resistor stability will be the subject of future papers. However, we will touch on the subject of the $1$~G$\Omega$ resistor stability again in section VI.C, as this was an important limiting factor in performing electron pump measurements with relative uncertainties much below $1$~ppm.

\section{independent tests}

\begin{figure}
\includegraphics[width=8.5cm]{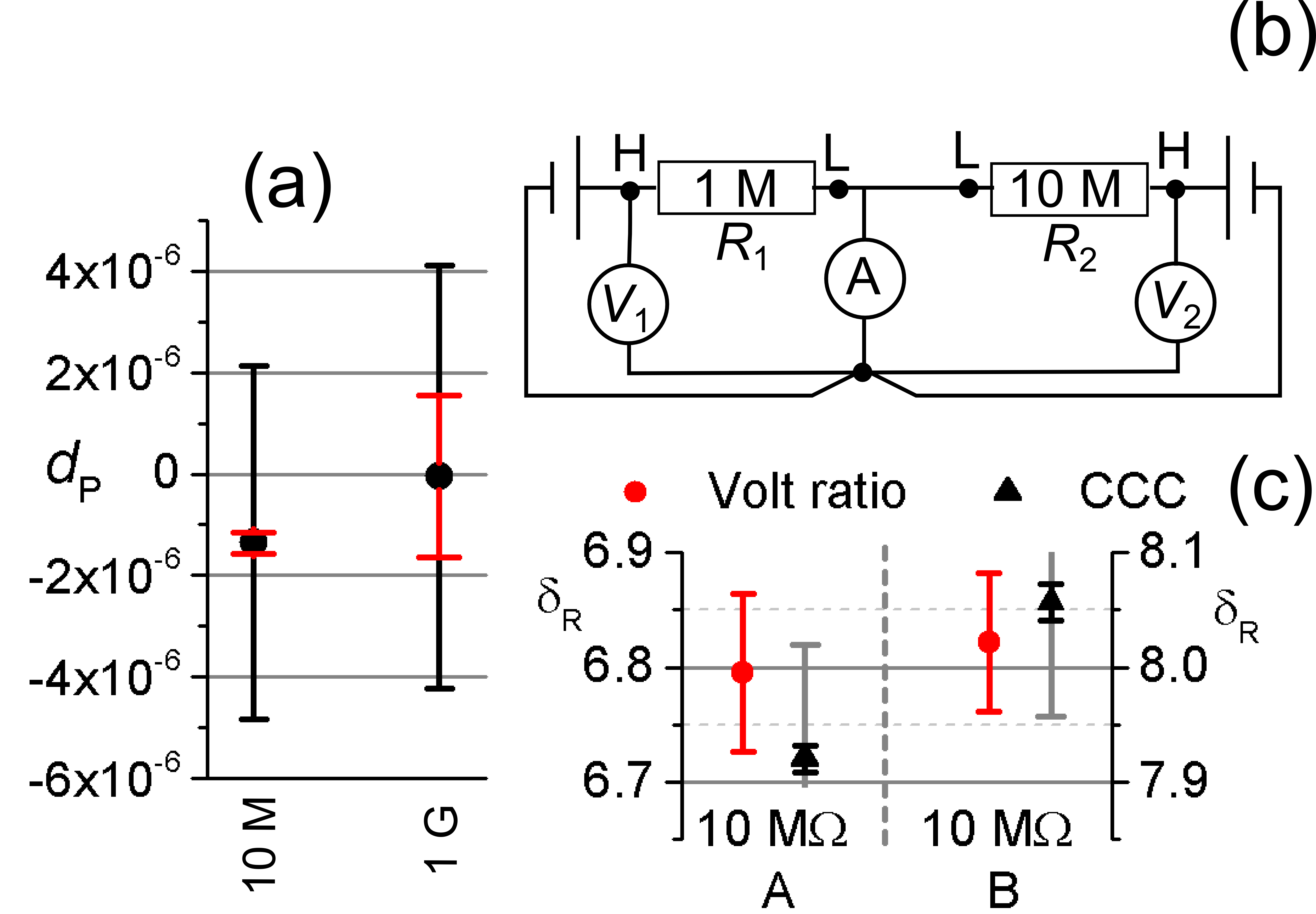}
\caption{\label{VoltRatioFig}\textsf{(a): degree of equivalence $d_{\text{P}}$ and associated uncertainty (large black error bars) for NPL participation in the euromet EM-K2 comparison of $10$~M$\Omega$ and $1$~G$\Omega$ resistors. The NPL CMC uncertainty is shown as the smaller red error bars. Both error bars show expanded $k=2$ uncertainties. (b): schematic circuit diagram of a voltage ratio bridge used to measure a $1$~M$\Omega:10$~M$\Omega$ resistance ratio. (c): Comparison of the voltage ratio bridge measurements with CCC measurements on two $10$~M$\Omega$ standards, where $R_{\text{10M}}=10$~M$\Omega(1+\delta_{\text{R}}/10^6)$. The left (right) pair of data points refer to the left (right) y-axis. Error bars for the voltage ratio bridge measurement indicate the total $1\sigma$ uncertainty. The black error bars on the CCC data points indicate the type A uncertainty, and the grey error bars indicate the total CMC uncertainty.}}
\end{figure}

Before embarking on a detailed evaluation of the bridge uncertainty, we briefly consider four cases where the bridge has been compared with other measurement methods. Firstly, the most recent relevant intercomparison, conducted in 2005-2007, was Euromet EM-K2, of $10$~M$\Omega$ and $1$~G$\Omega$ resistors \cite{jeckelmann2010final}. The degrees of equivalence and their associated $k=2$ uncertainty are plotted in figure ~\ref{VoltRatioFig} (a). Also plotted are the NPL CMC uncertainties. It can be seen that the comparison uncertainty was considerably larger than the CMC uncertainties for both the resistor values. This was mainly due to a large transport uncertainty component $\sim 2$~ppm contributing to the uncertainty in the degree of equivalence \cite{jeckelmann2010final}. While the comparison did not find any gross errors in the CCC measurements, it did not test the CCC at its design uncertainty level.

Secondly, the bridge was compared with a voltage ratio technique for measuring a $1$~M$\Omega:10$~M$\Omega$ resistance ratio. A schematic diagram of the voltage ratio bridge is shown in figure~\ref{VoltRatioFig} (b). Two voltage sources applied $1$~V and $-10$~V to the two-terminal $1$~M$\Omega$ and $10$~M$\Omega$ resistors respectively, and the voltages were measured by two precision DVMs (HP3458A). A sensitive ammeter connected to the junction of the low current resistor terminals measured the residual error current, and the value of the unknown resistor was given by $R_{\text{2}} = R_{\text{1}}[-\Delta V_{\text{2}}/(R_{\text{1}}\Delta I + \Delta V_{\text{1}})]$, where $\Delta V_{\text{1}}$,$\Delta V_{\text{2}}$ and $\Delta I$ are the difference signals recorded by the two voltmeters and the ammeter when the signs of the voltages were reversed. The link between the low current resistor terminals was made with a thick braid, $10$~cm long, and the ammeter was connected close to the $1$~M$\Omega$ terminal, roughly dividing the length of the braid in the ratio $1:9$. The voltmeters were calibrated directly against a Josephson array at least once every hour, and the maximum relative type B uncertainty due to the voltage measurement was $\sim 10^{-8}$. In fact, the voltage ratio method offers a comparable uncertainty to the CCC, but is impractical for routine use due to the requirement for frequent calibration of two voltmeters. As can be seen in figure~\ref{VoltRatioFig} (c), the voltage ratio bridge gave results in agreement with the CCC, within the combined uncertainties of the two measurements. 

Thirdly, an ULCA \cite{drung2015ultrastable} calibrated at PTB with a relative uncertainty of $\sim 0.1$~ppm was transported to NPL and LNE, with the objective of testing its gain stability under transportation\cite{drung2015validation}. At NPL, the ULCA was used to measure $I_{\text{R}}$ from the reference current generator employed in the electron pump measurements. Both $100$~M$\Omega$ and $1$~G$\Omega$ resistors were used to generate the reference current, and in both cases agreement of $\sim 0.5$~ppm was obtained with the ULCA indicated current, based on the PTB calibration. Since the study of Ref.\cite{drung2015validation}, a growing number of ULCA transfers now indicate that the ULCA gain is stable at at least the part in $10^6$ level under transportation, and the results of Ref. \cite{drung2015validation} can be interpreted as checking the NPL traceability to $100$~M$\Omega$ and $1$~G$\Omega$. In a fourth series of experiments, to be presented in an upcoming paper\cite{giblin2018interlaboratory}, an ULCA was calibrated at NPL and used to measure a $1$~G$\Omega$ resistor. Comparison between the ULCA measurement and a CCC measurement of the same resistor produced agreement better than 1 part in $10^{7}$. 

Collectively, these four tests rule out gross errors in the CCC and support the NPL CMC relative uncertainties (k=2) of $2 \times 10^{-7}$, $4 \times 10^{-7}$ and $1.6 \times 10^{-6}$ for $10$~M$\Omega$, $100$~M$\Omega$ and $1$~G$\Omega$ respectively. They provide a base-line from which we commence our detailed evaluation of the CCC uncertainty.

\section{CCC uncertainty evaluation}

Now we turn to the main aim of this paper: an evaluation of the type B uncertainty in the measurement of the ratio $R_{\text{1}}/R_{\text{2}}$ with the CCC operated as a passive current divider without a current servo. This is due to four contributions: leakage resistance, SQUID non-linearity, the correction due to series resistances $r_{\text{1}}$ and $r_{\text{2}}$ and finally the winding ratio error which is evaluated using $1:1$ ratio error tests (RATs).

\subsection{leakage resistance}

Referring to figure \ref{BridgeDiagramFig}, the bridge is sensitive to leakage at points A and B, in between the resistors and the CCC windings. As already noted, point A is maintained at screen potential using a Wagner network, which minimises the effect of a leakage resistance between this point and screen. However, the lead resistances between points A and B, and the low terminals of $R_{\text{1}}$ and $R_{\text{2}}$ respectively, result in a small voltage between the low resistor terminals and screen which will drive a current in the leakage resistance distributed between the resistor terminals and leads. To asses the magnitude of leakage effects, the 'low' side of the bridge circuit was isolated by disconnecting it from the remainder of the bridge at points L, C and D. The Wagner amplifier was also disconnected at point A, and the detector was disconnected inside the CCC electronics enclosure (the detector itself has a differential input resistance $>1$~T$\Omega$; it was disconnected to avoid possible damage to its sensitive input circuit due to the test voltage). Thus, the leakage measurements include the CCC windings, $N_{\text{1}}$ winding selection relays, low side resistor connecting cables and the resistors themselves. A total of $8$ measurements were made: leakage from point A to screen with $N_{\text{1}} = 100,1000,10000$, leakage from point B to screen with $N_{\text{2}} = 10000,10001$, and leakage from point A to point B with $N_{\text{1}} = 100,1000,10000$. In all cases, the leakage resistance was measured by applying a test voltage of $\pm 10$~V and measuring the difference current using a source-measure unit (Keithley 6430), which in all cases was less than $20$~pA, setting a lower limit to the leakage resistances $R_{\text{L}}$ of $1$~T$\Omega$. Leakage currents cause errors $\sim r/R_{\text{L}}$, where $r$ is the lead resistance\cite{williams2010automated}. Since winding and lead resistances are roughly $10$~$\Omega$, we conclude that the effect of leakages on the relative uncertainty in measuring $R_{\text{1}}/R_{\text{2}}$ is at the $10^{-11}$ level, completely negligible compared to other uncertainty components.

\subsection{Linearity of SQUID readout}

\begin{figure}
\includegraphics[width=8.5cm]{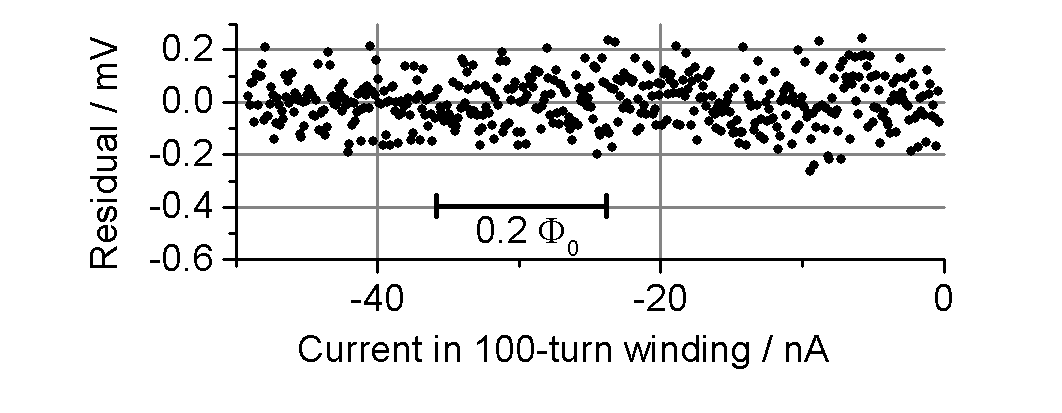}
\caption{\label{SquidLinFig}\textsf{Residual of a linear fit to the SQUID output voltage $V_{\text{S}}$ as a function of the current in the $100$-turn winding.}}
\end{figure}

The CCC employs a commercial SQUID and associated feedback electronics (Quantum Design model 550). The SQUID difference readings $\Delta V_{\text{S}}$ and $\Delta V_{\text{S}}'$ are used directly to calculate the resistance ratio, so the linearity of the SQUID must be considered as a contribution to the uncertainty. Referring to figure \ref{BridgeDiagramFig}, possible errors due to SQUID non-linearity were evaluated by removing $R_{\text{2}}$ and measuring the SQUID output $V_{\text{S}}$ while sweeping $V_{\text{SOURCE}}$ through a small range (such that $N_{\text{1}}I_{\text{1}}/S_{\text{CCC}} < 1 \Phi_{0}$). $V_{\text{SOURCE}}$ was measured using an 8.5-digit voltmeter with better than $10^{-6}$ linearity. A linear fit was performed to $V_{\text{S}}$ as a function of $I_{\text{1}}$. An example data set is shown in figure \ref{SquidLinFig}, which shows the fit residuals as a function of $I_{\text{1}}$. There is no structure visible in the residuals indicative of non-linearity within the standard deviation ($\sim 0.1$~mV) of the points, and we conservatively interpret this standard deviation as setting an upper limit to non-linearity errors over a $0.5$~V range of SQUID feedback voltage. Based on this data, we can assign an approximate uncertainty in the squid voltage due to possible non-linearity, of $10^{-4}/0.5 = 2\times10^{-4} \Delta V_{\text{S}}$. The relative uncertainty in the resistance ratio is then $2\times10^{-4} \Delta V_{\text{S}}/kN_{\text{1}}I_{\text{1}}$. For a typical case of a $100$~M$\Omega$ calibration at $100$~V (figure \ref{RawDataFig} (a)), $\Delta V_{\text{S}} \sim 0.1$~V and the non-linearity uncertainty is $< 1 \times 10^{-8}$. Similarly, for the $1$~G$\Omega$ raw data shown in figure \ref{RawDataFig} (b), $\Delta V_{\text{S}} \approx 3$~mV, giving a non-linearity uncertainty of $8\times 10^{-9}$. However, for the specific case of the SET $1$~G$\Omega$ this is clearly an over-estimate for the fortuitous reason that $\Delta V_{\text{S}} \approx -\Delta V_{\text{S}}'$, a consequence of the resistor deviating from its nominal value by $\approx 5\times 10^{-5}$. The two parts of the calibration, with $N_{\text{2}}=10000$ and $N_{\text{2}}=10001$, therefore sample the same part of the SQUID $V-\Phi$ characteristic, and the linearity error becomes less than $10^{-9}$.

\subsection{series resistance correction}

The CCC source voltage $V_{\text{SOURCE}}$ and the detector voltage $V_{\text{D}}$ were both calibrated with a relative uncertainty of $10^{-2}$ using calibrated voltage sources and DMMs. For calibrations of the SET $1$~G$\Omega$, $\Delta V_{\text{D}}/2V_{\text{S}}=5 \times 10^{-7}$ with a relative uncertainty of $0.014$, so the uncertainty in this quantity contributes a relative uncertainty of $7 \times 10^{-9}$ to the uncertainty in $R_{\text{2}}/R_{\text{1}}$ using equation (4). As an additional check, the value of $r_{\text{1}}$ obtained from $\Delta V_{\text{D}}$ data and equation (3) (after measuring $r_{\text{2}}$ with modest precision using a hand-held multimeter) was compared with the value measured directly by connecting a 4-wire DMM to appropriate circuit test points. The agreement was better than $0.05$~$\Omega$.

\subsection{winding ratio error}

The basic configuration of the ratio error test (RAT) is well known: the same current is passed through two windings with equal numbers of turns, $N_{\text{1}} = N_{\text{2}} = N$ but opposite sense so that zero flux will be measured by a perfect comparator. The current is periodically reversed, giving a difference in the excitation current $\Delta I_{\text{1,2}}$. The squid signal is recorded and the SQUID difference voltage $\Delta V_{\text{S}}$ is extracted. The error in the turns ratio is given by $\delta_{\text{N}} \equiv \frac{\Delta V_{\text{S}}}{kN\Delta_{\text{I}}}$. RATs have historically been used mainly for two purposes: firstly to check for gross errors in a new comparator such as cracks in the shield, or incorrect numbers of turns, and secondly to evaluate small errors due to flux penetration into the comparator associated with the finite number of CCC shield overlaps. Recent work\cite{drung2015improving} further identified a 'low-flux' regime in which a finite SQUID difference signal could be obtained in a RAT due to non-linear rectification of noise in the SQUID itself, effectively setting a limit of $\sim 10^{-6} \Phi _{0}$ to the resolution of a CCC.

A full binary build-up was performed on our CCC prior to its commissioning, and no errors were found in Ratio error tests (RATs) at the $10^{-8}$ level. However, the configuration of the electronics, connecting cables and CCC down-leads employed for routine use only allows access to a sub-set of the windings, and a repeat of the full build-up was not performed in this study. Instead, we focused on sensitive RATs with $N=1$, $2$, and $10000$ turns. We performed sensitive RATs using two different current sources: the bridge's internal voltage source in series with a $10$~M$\Omega$ standard resistor for generating currents up to $10$~$\mu$A, and a commercial current source (Yokogawa GS200) with a single-pole $30$~Hz RC filter on the output for larger currents. RATs were performed with the current sources connected with both polarities, where `polarity' refers to the sense of the physical connection between the current source and the windings. Altogether, the RATs spanned a range of $10$~mAturns $\leq N\Delta I_{\text{1,2}} \leq 2$~Aturns. At the low limit of this range, $N\Delta I_{\text{1,2}}/S_{\text{CCC}}=1.8$~$\mu\Phi_{0}$, close to the onset of non-linear rectification effects\cite{drung2015improving}.

\begin{figure}
\includegraphics[width=9cm]{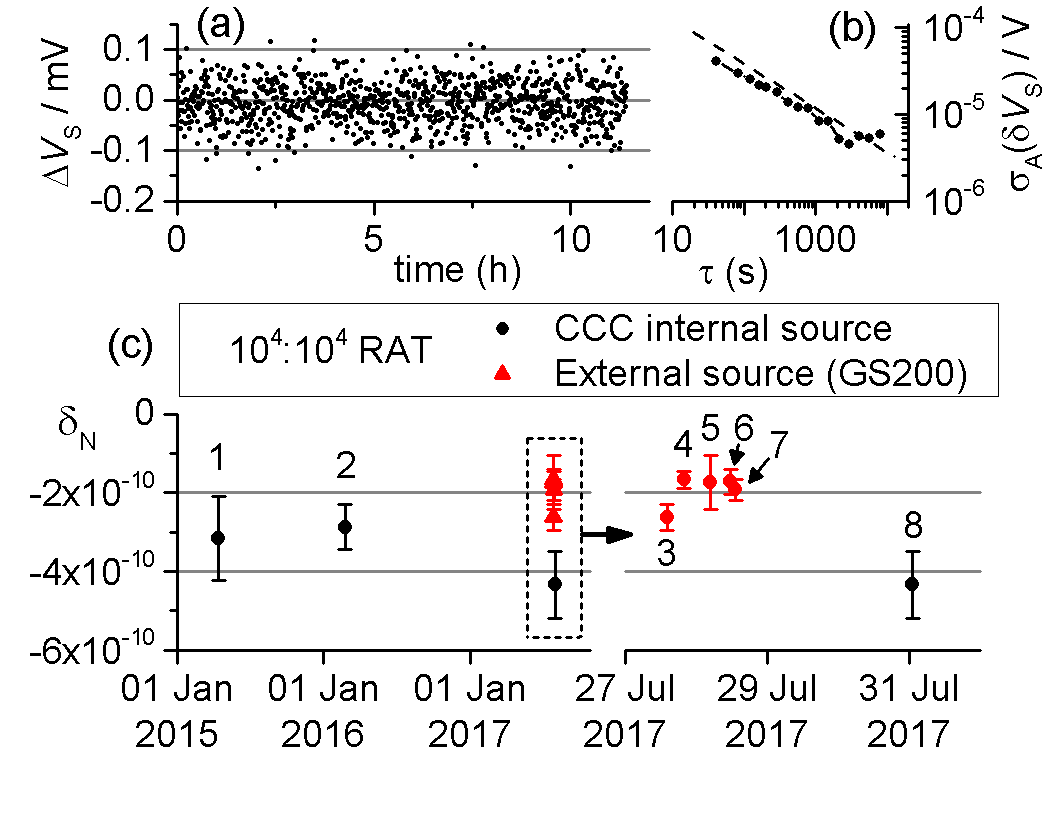}
\caption{\label{10000TurnsErrFig}\textsf{(a): Squid difference voltage for a $10^4$-turn RAT with $\pm 10$~$\mu$A. Each data point is calculated from $32$ seconds of raw data; $16$ seconds with each current polarity. (b): Allan deviation of the data plotted in (a). The dashed line is a guide to the eye with gradient $1/\sqrt{\tau}$. (c): Ratio error $\delta_{\text{N}}$ for $8$ RATs. Error bars are the statistical uncertainty, evaluated as the standard error on the mean. The right panel shows RATs $3$-$8$ on an expanded time axis.}}
\end{figure}

Focusing first on the high-turns RATs, we performed a total of 8 RATs, with parameters given in table I, and a current reversal every $16$~s. The individual $\Delta V_{\text{S}}$ measurements for one test are plotted in figure \ref{10000TurnsErrFig}(a), with the Allan deviation of these data in figure \ref{10000TurnsErrFig}(b). The mean of these measurements is $\Delta V_{\text{S}} = -6.6$~$\mu$V, with a standard error on the mean of $1.4$~$\mu$V. This corresponds to a finite ratio error of $\delta_{\text{N}} = -2.9\times 10^{-10}$, which is qualitatively consistent with the number of shield overlaps (3) in the CCC construction. Using the same test setup, consistent results were obtained in 3 RATs spread over 3 years (RATs 1,2 and 8). Additional 10000-turn RATs (3-7) were performed using the commercial current source with currents up to $100$~$\mu$A, and these showed that there is no marked dependence of $\delta_{\text{N}}$ on $\Delta I_{\text{1,2}}$ or the current source polarity. We conclude that the high-turns RATs are measuring the geometric error due to the finite CCC shield overlap, with a relatively small contribution from non-linear rectification errors. 

\begin{figure}
\includegraphics[width=9cm]{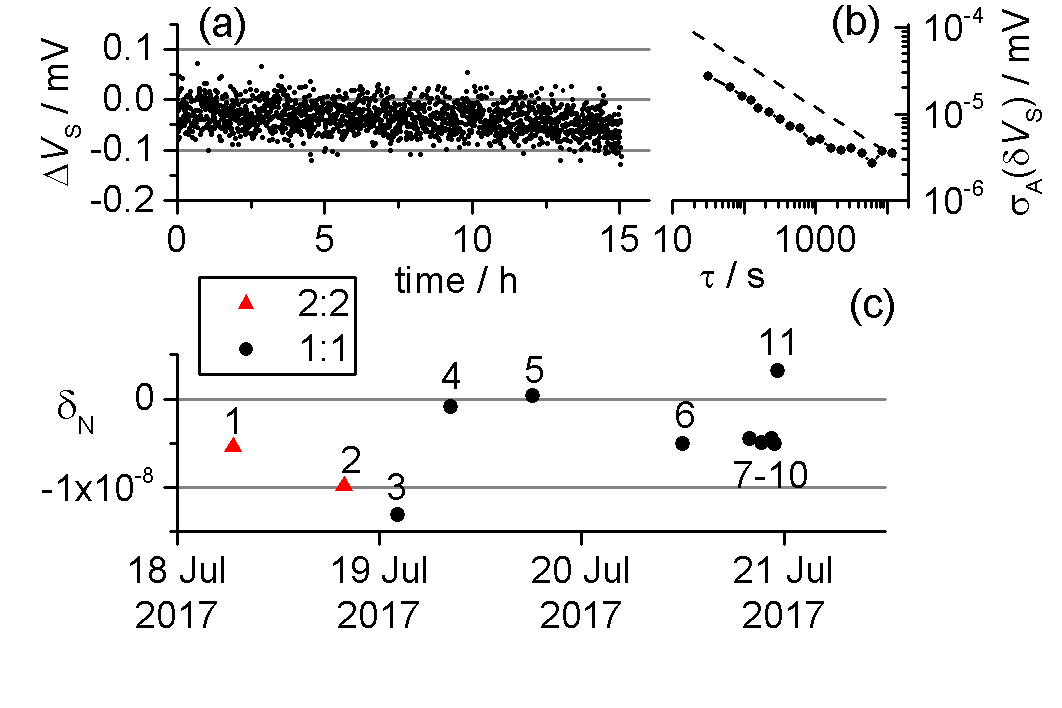}
\caption{\label{1TurnErrFig}\textsf{(a): Squid difference voltage for a 2-turn RAT with $\pm 10$~mA. Each data point is calculated from $32$ seconds of raw data; $16$ seconds with each current polarity. (b): Allan deviation of the data plotted in (a). The dashed line is a guide to the eye with gradient $1/\sqrt{\tau}$. (c): Ratio error $\delta_{\text{N}}$ for $11$ RATs. Tests $1$ and $2$ respectively were evaluated from the first and last $500$ cycles of the data shown in panel (a). The statistical uncertainty evaluated as the standard error on the mean is typically the same size as the data points.}}
\end{figure}

In contrast, the 1- and 2-turn RATs (figure \ref{1TurnErrFig} and table II) show a much larger ratio error, as large as $\delta_{\text{N}} \sim 10^{-8}$. Moreover, $\delta_{\text{N}}$ was seen to fluctuate on time-scales longer than a few hours, as clearly evidenced by the data set of figure \ref{1TurnErrFig}(a). The beginning and end of this data set is analysed to yield RAT measurements 1 and 2 (figure \ref{1TurnErrFig}(c) and table 2) which are clearly different within their statistical uncertainties. Considering all the low-turn RATs, while $\delta_{\text{N}}$ does not correlate with $\delta_{\text{I}}$, there is a clear influence of current polarity: RATs 4, 5 and 11, which employed reverse polarity, yielded much smaller $\delta_{\text{N}}$ than the other RATs. We are led to conclude that non-linear rectification effects have a bigger effect for these low-turn RATs than the high-turn RATs This is not an unexpected discovery, because the low-turn RATs employ a lower overall ampere-turn linkage in the comparator. For the smallest current used (tests 5 and 6), $\delta_{\text{I}}=10$~mA, and an error $\delta_{\text{N}} = 10^{-9}$ corresponds to a SQUID flux difference of $1.8$~$\mu\Phi_{0}$. This is close to the threshold where errors in RATs have been associated with rectification effects\cite{drung2015improving}. A further possibility is that the 1- and 2-turn RATs are probing the geometrical error, but the single turns are much more sensitive to the geometry of the flux penetration than the large-turn windings. If the single turns were mechanically unstable, fluctuating results could be obtained in RATs.

Because we were not able to perform a full build-up, we cannot quantitatively assess the error associated with the turns ratio $N_{\text{2}}/N_{\text{1}}$ that enters equation (4). However, we can set an upper limit to possible turns-ratio errors by using the low-turns RATs (figure \ref{1TurnErrFig} (c)) as representing a worst-case. From this, we assign a relative uncertainty contribution of $1 \times 10^{-8}$ to the ratio $R_{\text{2}}/R_{\text{1}}$.

\begin{table}
\caption{Parameters for $10^4$-turn RATs}
\centering
\setlength{\tabcolsep}{8pt}
\begin{tabular}{c c c c c}
\hline\hline
Test & Current source & Current ($\mu$A) & Polarity & \\[0.5ex]
\hline
1 & Internal & $\pm 10$ & F \\
2 & Internal & $\pm 10$ & F \\
3 & GS200 & $\pm 50$ & F \\
4 & GS200 & $\pm 50$ & R \\
5 & GS200 & $\pm 10$ & R \\
6 & GS200 & $\pm 100$ & R  \\
7 & GS200 & $\pm 100$ & F \\
8 & Internal & $\pm 10$ & F \\ [1ex]
\hline
\end{tabular}
\end{table}

\begin{table}
\caption{Parameters for $1$- and $2$-turn RATs}
\centering
\setlength{\tabcolsep}{8pt}
\begin{tabular}{c c c c c}
\hline\hline
Test & Turns & Current (mA) & Polarity & \\[0.5ex]
\hline
1 & 2 & $\pm 10$ & F \\
2 & 2 & $\pm 10$ & F \\
3 & 1 & $\pm 10$ & F \\
4 & 1 & $\pm 10$ & R \\
5 & 1 & $\pm 5$ & R \\
6 & 1 & $\pm 5$ & F  \\
7 & 1 & $\pm 10$ & F \\
8 & 1 & $\pm 10$ & F \\
9 & 1 & $\pm 50$ & F \\
10 &1 & $\pm 100$ & F \\ 
11 & 1 & $\pm 100$ & R \\ [1ex]
\hline
\end{tabular}
\end{table}

\section{Additional considerations}

Up until now we have evaluated contributions to the uncertainty in the measured resistance ratio $R_{\text{1}}/R_{\text{2}}$ due to the CCC instrument. It remains to consider the uncertainty in the known $10$~M$\Omega$ resistor $R_{\text{1}}$. For measurements of the SET $1$~G$\Omega$, the voltage coefficient and time dependence were also significant contributions to the combined uncertainty, and these are also discussed below. 

\subsection{Reference resistor}

For the recently reported electron pump measurements\cite{zhao2017thermal}, the $1 \sigma$ uncertainty in the $10$~M$\Omega$ reference resistor was taken simply to be the CMC relative uncertainty, $1 \times 10^{-7}$. 

\subsection{Voltage coefficient}

\begin{figure}
\includegraphics[width=9cm]{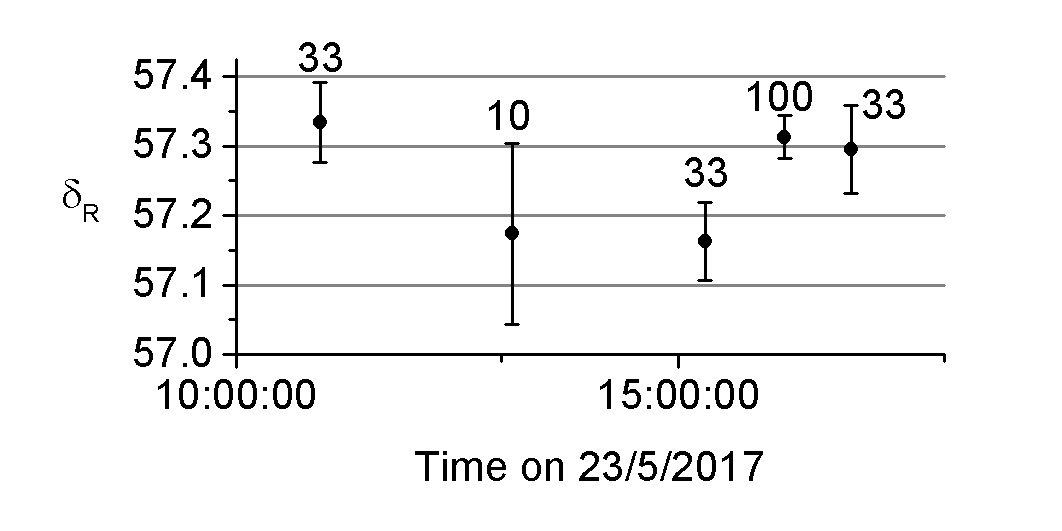}
\caption{\label{VoltCoeffFig}\textsf{Values of the SET $1$~G$\Omega$ obtained from CCC measurements at different voltages. $R_{\text{1G}}=1$~G$\Omega(1+\delta_{\text{R}}/10^6)$. The voltage is indicated next to each data point, and the error bars show the type A uncertainty.}}
\end{figure}

For the electron pump measurements, possible voltage co-efficients in the $1$~G$\Omega$ resistor must be considered. The resistor is calibrated with an excitation voltage of $\pm 33$~V, but measurements of electron pump current $I_{\text{P}} \approx 160$~pA require a much smaller voltage, $\approx 160$~mV across the resistor. In evaluating the uncertainty due to possible voltage coefficients we take a conservative approach, and attempt to set an upper limit to possible voltage effects by measurement of the resistor over a range of voltages. Earlier CCC measurements\cite{giblin2012towards} at $100$~V and $10$~V ruled out relative voltage co-efficients at the level of $\sim 5 \times 10^{-7}$. In figure \ref{VoltCoeffFig} we present a recent set of $1$~G$\Omega$ measurements with different excitation voltages. No systematic voltage-dependent effect can be discerned in this data, although the poor signal-to-noise ratio at low voltages combined with time-dependent drifts in the resistance\cite{giblin2018limitations} make it difficult to eliminate the possibility of voltage-dependence at the $10^{-7}$ level on the basis of this data. The ULCA \cite{drung2015ultrastable} provided another method for low-voltage measurements of $1$~G$\Omega$ resistors, because the resistor can be compared directly with the ULCA trans-resistance gain which is also nominally $1$~G$\Omega$. ULCA measurements with $\pm 4.8$~V excitation and CCC measurements with $\pm 33$~V agreed within relative combined uncertainties of $\sim 10^{-7}$ in three sets of measurements made on different days\cite{giblin2018interlaboratory}. The uncertainty in the comparison was limited by the need to perform the measurements within a short time-scale to avoid resistor drift effects. Based on this data set, we assign a relative uncertainty due to possible voltage co-efficients of $10^{-7}/ \sqrt{3} = 5.8 \times10^{-8}$. A future measurement strategy for the SET $1$~G$\Omega$ could be to measure it using the ULCA at low voltage, and assume that there is no further voltage co-efficient between the ULCA measurement voltage ($\sim 5$~V) and the SET voltage ($\sim 0.16$~V). Alternatively, the ULCA could be used to measure the electron pump current directly\cite{stein2015validation,stein2016robustness}, a possibility further discussed in section VII.

\subsection{Resistor drift}

\begin{figure}
\includegraphics[width=9cm]{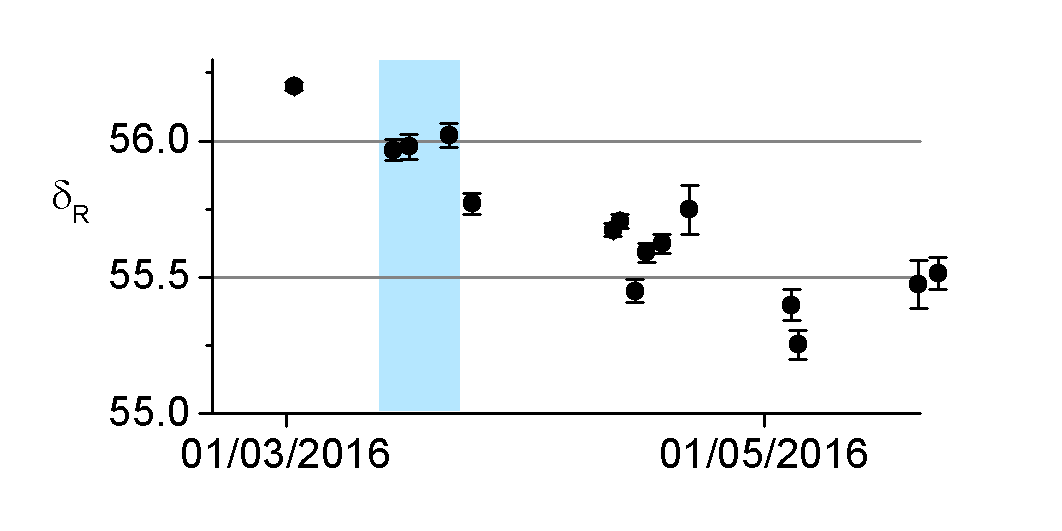}
\caption{\label{1GDriftFig}\textsf{Values of the SET $1$~G$\Omega$ obtained from CCC measurements during a measurement campaign on a silicon electron pump\cite{zhao2017thermal}, where $R_{\text{1G}}=1$~G$\Omega(1+\delta_{\text{R}}/10^6)$. The shaded area indicates the time during which the highest-accuracy pump measurements were performed.}}
\end{figure}

As already mentioned, standard resistors of $100$~M$\Omega$ and $1$~G$\Omega$ based on thick-film resistive elements exhibited instability on time-scales of hours to days. This instability was clearly visible as a flattening (characteristic of $1/f$ noise) of the Allan deviation of the measured resistance on time-scales longer than $1$~hour, and detailed data will be presented elsewhere\cite{giblin2018limitations}. Here, in figure \ref{1GDriftFig}, we present a plot of the measured values of the SET $1$~G$\Omega$ over a time-base of several months, covering the electron pump measurement campaign reported in\cite{zhao2017thermal}. During the most intensive periods of pump measurements, the resistor was calibrated every $1-3$ days. No assumptions were made about the behavior of the resistor in between calibrations (for example, it was not assumed that it exhibited linear drift), so its value during the pump measurement was taken to be $(R_{\text{after}} + R_{\text{before}})/2$ with standard uncertainty obtained from a square distribution as $|R_{\text{after}} - R_{\text{before}}|/2\sqrt{3}$. Here, $R_{\text{before}}$ and $R_{\text{after}}$ are the values of the resistor before and after the pump measurement respectively. The two key measurement runs reported in\cite{zhao2017thermal} were made in the time period highlighted by the blue shading in figure \ref{1GDriftFig}, and the relative uncertainty components due to drift resistor drift were $1 \times 10^{-8}$ for both runs. Nevertheless, we did not feel that this fairly small number captured our uncertainty in knowing the resistor value in between calibrations, because as already noted, in some calibration runs lasting several hours, the Allan deviation exhibited a transition to $1/f$ noise. For this reason, we assigned a type A uncertainty in the resistor value based on the Allan deviation, and not the standard error on the mean, resulting in relative uncertainty contributions of $1.5 \times 10^{-7}$ and $7 \times 10^{-8}$ for the two key runs reported in\cite{zhao2017thermal}. At present, the stability of high-value resistors is still an area of investigation, extrapolation of linear drift lines is not supported by the available data (such as figure \ref{1GDriftFig}) and each case needs to be treated individually.

\section{discussion}

In table III we present a summary of the relative type B uncertainty contributions ($1\sigma$) evaluated in this paper, ordered by decreasing size, for the case of a $1$~G$\Omega$ resistor used at low voltage to generate a reference current for a single-electron experiment. We have taken as our starting point the $1\times 10^{-7}$ CMC uncertainty for the $10$~M$\Omega$ reference resistor, and this is by far the largest contributor to the relative combined uncertainty of $1.17\times 10^{-7}$. The second-largest term comes from empirical evaluation of the voltage co-efficient of the resistor. As noted in section VI.B, our ability to evaluate this term is limited by the long averaging times required to measure the resistor at low voltage, combined with instability in the resistor itself. The remaining terms contribute an insignificant amount to the combined uncertainty. In a practical electron pump measurement, we must add to the combined uncertainty in table III some significant additional terms\cite{zhao2017thermal} related to the voltmeter calibration and drift between calibrations ($\sim 5 \times 10^{-8}$), the type A uncertainty in the resistor calibration ($\sim 1 \times 10^{-7}$) and the type A uncertainty in the measurement of $I_{\text{P}}$ itself ($\sim 2 \times 10^{-7}$ for a typical experimental run lasting 10-15 hours). This last, largest, term arises from the Johnson-Nyquist current noise $=\sqrt{4 k_{\text{B}}T/R} \approx 4$~fA$/\sqrt{Hz}$ in the $1$~G$\Omega$ resistor, and we note that it could be reduced by a factor of $\sqrt{3}$ by using a standard ULCA to measure $I_{\text{P}}$ directly.

The data in table III are for the specific case relevant to the SET $1$~G$\Omega$. For routine calibrations of $100$~M$\Omega$ and $1$~G$\Omega$, the voltage co-efficient term can be neglected because calibration values are quoted at the calibration voltage, usually $100$~V. The SQUID linearity term may be as large as a few parts in $10^8$ for resistors with a large deviation from nominal, but to a good approximation the $1\sigma$ relative standard uncertainty is $\approx 1 \times 10^{-7}$, dominated by the uncertainty in the $10$~M$\Omega$ reference resistor. This re-evaluated uncertainty is smaller than the CMC uncertainties by a factor $2$ and $8$ for $100$~M$\Omega$ and $1$~G$\Omega$ resistors respectively. The reason for the large CMC uncertainties is that loop-closure tests played a major role in the original evaluation of the bridge performance under routine calibration conditions. In these tests, for example, a $1$~G$\Omega$ resistor would be measured against a $10$~M$\Omega$ reference resistor in a single $100:1$ steps, and two $10:1$ steps using an intermediate $100$~M$\Omega$ standard. These tests can be re-evaluated in the light of recent findings of the short-term instability of high-value standard resistors. Discrepancies in the original $1$~G$\Omega$ loop-closure tests were most likely caused by drifts in the resistors on the time-scales of a few hours required to perform the measurements. For this reason, the CMC uncertainties are an accurate representation of the realistically achievable uncertainty in transferring a resistance value to a customer's laboratory, although they are a pessimistic expression of the CCC performance.

The question naturally arises as to whether the CMC uncertainty in the $10$~M$\Omega$ reference resistor is over-estimated, and whether an extension of the investigations of this paper to $10$~M$\Omega$ and lower resistance values would yield a profitable reduction in uncertainty for SET measurements. The problem is that reduction of type B uncertainty terms leaves the stability of the $1$~G$\Omega$ resistor (figure \ref{1GDriftFig}) increasingly exposed as the most problematic factor in the SET experiment. The requirement for almost daily calibration of the resistor interrupts the measurements and makes it difficult to perform long continuous measurements, for example, to evaluate the flatness of an electron pump plateau. Unless more stable high-value resistors can be found, using the ULCA to measure $I_{\text{P}}$ directly appears to be the best route. The drift in the trans-resistance gain of the ULCA is smaller, and more predictable\cite{drung2015ultrastable}, than the drift of a $1$~G$\Omega$ standard resistor. Furthermore, variants of the ULCA have been tested with larger resistances in the input current-scaling network\cite{drung2017ultrastable}, offering even lower Johnson-Nyquist current noise than the standard ULCA. It is conceivable that a measurement of the present generation of electron pumps, with $I_{\text{P}} \sim 160$~pA, could be made with a relative uncertainty of $5 \times 10^{-8}$. Reduction of the uncertainty significantly below this level requires more frequent calibration of the ULCA, implying shorter averaging times, and higher pumps currents, such as may be offered by pumps based on single traps in silicon\cite{yamahata2017high}.

\section{conclusions}

We conclude that $100$~M$\Omega$ and $1$~G$\Omega$ resistors can be measured using the NPL high-resistance CCC with type B uncertainties not significantly larger than $1\times 10^{-7}$. This capability has allowed the investigation of prototype single-electron current standards with combined uncertainties dominated by the type A component, about $2$ parts in $10^7$ for $160$~pA averaged over $10$ hours. 

\begin{table}
\caption{Relative type B uncertainty contributions for measuring $1$~G$\Omega$}
\centering
\setlength{\tabcolsep}{8pt}
\begin{tabular}{l c c}
\hline\hline
Component & Section reference & Uncertainty \\[0.5ex]
\hline
10 M$\Omega$ reference & I (CMC) & $1 \times 10^{-7}$  \\
Voltage coefficient & VI.B & $5.8 \times 10^{-8}$ \\
Turns ratio & V.D & $1 \times 10^{-8}$ \\
Series resistance & V.C & $7 \times 10^{-9}$ \\
Squid linearity & V.B & $1 \times 10^{-9}$ \\
leakage resistance & V.A & $1 \times 10^{-11}$ \\ 
\hline
Total &    & $1.16 \times 10^{-7}$ \\[1ex]
\hline
\end{tabular}
\end{table}

\begin{acknowledgments}
This research was supported by the UK department for Business, Energy and Industrial Strategy, the Joint Research Project 'Quantum Ampere' (JRP SIB07) within the European Metrology Research Programme (EMRP) and the EMPIR Joint Research Project 'e-SI-Amp' (15SIB08). The The EMRP is jointly funded by the EMRP participating countries within EURAMET and the European Union. The European Metrology Programme for Innovation and Research (EMPIR) is co-financed by the Participating States and from the European Union's Horizon 2020 research and innovation programme. The author would like to thank Jonathan Williams, Nick Fletcher, Dietmar Drung and Martin Goetz for stimulating discussions.
\end{acknowledgments}

\bibliography{SPGrefs_CCCPaper}

\end{document}